\documentclass[reprint,showpacs,amsmath,amssymb,xcolor=dvipsnames,prl,longbibliography,superscriptaddress,floatfix]{revtex4-2}
\pdfoutput=1

\usepackage[pdftex]{graphicx}
\usepackage{bm}
\usepackage[pdftex,colorlinks=true,citecolor=blue,urlcolor=blue,linkcolor=black]{hyperref}
\usepackage{braket}
\usepackage[dvipsnames]{xcolor}
\usepackage{siunitx}
\usepackage{ mathrsfs }
\usepackage{lipsum}  
\usepackage{comment}

\newcommand{\revision}[1]{\textcolor{black}{#1}}

\begin{document}

\footnotetext[1]{Note that choosing a larger trap frequency $\omega \propto \sqrt{P}W^{-2}$ over a larger waist reduces the laser power required, as long as condition \eqref{eq: spatial extent} is fulfilled.}
\footnotetext[2]{Note that a different radial trap frequency was used in the non-interacting regime. The position and momenta are thus scaled to the natural units $l_\text{HO,r}$ and $p_\text{HO} = \sqrt{\hbar m\omega_\text{OT,r}}$ of the initial trap.}

\title{Magnifying the Wave Function of Interacting Fermionic Atoms}
\author{Sandra Brandstetter}
    \thanks{These authors contributed equally to this work.}
    \affiliation{Physikalisches Institut der Universit\"at Heidelberg, Im Neuenheimer Feld 226, 69120 Heidelberg, Germany}
\author{Carl Heintze}
    \thanks{These authors contributed equally to this work.}
    \affiliation{Physikalisches Institut der Universit\"at Heidelberg, Im Neuenheimer Feld 226, 69120 Heidelberg, Germany}
\author{Paul Hill}
    \affiliation{Physikalisches Institut der Universit\"at Heidelberg, Im Neuenheimer Feld 226, 69120 Heidelberg, Germany}
\author{Philipp M. Preiss}
    \affiliation{Physikalisches Institut der Universit\"at Heidelberg, Im Neuenheimer Feld 226, 69120 Heidelberg, Germany}
    \affiliation{Current adress: Max Planck Institute of Quantum Optics, Hans-Kopfermann-Str. 1, 85748 Garching, Germany}
\author{Maciej Ga\l ka}
    \affiliation{Physikalisches Institut der Universit\"at Heidelberg, Im Neuenheimer Feld 226, 69120 Heidelberg, Germany}
\author{Selim Jochim}
    \affiliation{Physikalisches Institut der Universit\"at Heidelberg, Im Neuenheimer Feld 226, 69120 Heidelberg, Germany}
\date{\today}

\begin{abstract}

Understanding many body systems is a key challenge in physics. Single atom resolved imaging techniques have unlocked access to microscopic correlations in ultracold quantum gases. However they cannot be used when the relevant length scales are obscured by the resolution of the detection technique. We present a matterwave magnification scheme, based on evolutions in optical potentials, tailored to magnify the wave function of atoms, such that all length scales can be resolved. To showcase this method, we image atoms in the strongly interacting regime, establishing a new way to characterize correlated systems.

\end{abstract}
\maketitle



Access to high-order correlation functions has become an essential tool in the quest to obtain a deeper understanding of strongly correlated many-body systems. To this end, a vast toolbox for the microscopy of ultracold atoms with single atom and spin resolution has been developed, allowing the in-depth exploration of both lattice systems -- using quantum gas microscopes~\cite{gross2021review} -- and continuous systems~\cite{bucker2009single, bergschneider2018spin, schellekens2005mcp,verstraten2025insitu}. This toolbox has for example been employed in the experimental microscopic observation of the superfluid-Mott insulator transition~\cite{bakr2010mott,sherson2010mott}, the demonstration of topological quantum states~\cite{lunt2024laughlin, Leonard2023} and quantum computing with neutral atoms~\cite{Bluvstein2023}. 

However, direct imaging is not possible in systems where the inter-particle spacing is on the order of the optical resolution. Conceptually, there are two ways to study such systems: either by using super-resolution techniques with sub-wavelength resolution~\cite{McDonald2019,Subhankar2019} or by magnifying the system prior to imaging in a way that preserves the correlations between constituents. \revision{This has been achieved using an accordion lattice~\cite{su2024fast} and by using matterwave optics techniques~\cite{shvarchuck2002bose,tung2010observation,murthy2014matter,asteria2021magnifier,Veit2021}. }

In this Letter we demonstrate the realization of a matterwave microscope for continuous systems in two dimensions. We magnify the wave function of few ultracold atoms by a factor of about 50, which  -- in combination with our established free space single atom and spin resolved imaging technique~\cite{bergschneider2018spin,holten2022cooper} -- enables access to correlations of arbitrary order. We characterize the performance of the matterwave microscope using the non-interacting harmonic oscillator ground state as a test target and demonstrate its capabilities by imaging a strongly interacting system with single atom and spin resolution in real space. 

Matterwave magnification is achieved by two subsequent rotations of the wave function in phase space ~\cite{murthy2014matter}, which are realized by a time evolution in two harmonic potentials, each described by $U_{1,2}=1/2m\omega_{1,2} ^2 x^2$. Here, $m$ is the mass of the particles, $\omega_{1,2}$ is the angular frequency of the respective harmonic oscillator and $x$ is the position in the two-dimensional plane to be imaged. In the Heisenberg picture, the time evolution of the single particle, real and momentum space operators -- $\hat{x}$ and $\hat{p}$, respectively -- in a harmonic potential are given by 
\begin{equation}
    \begin{bmatrix}
        \hat{x}(t)\\
        \hat{p}(t)
    \end{bmatrix} =
    \begin{bmatrix}
        \cos{\omega t} & \frac{1}{m\omega} \sin{\omega t}\\
        -m\omega\sin{\omega t} & \cos{\omega t}
    \end{bmatrix}
    \begin{bmatrix}
        \hat{x}(0)\\
        \hat{p}(0)
    \end{bmatrix}.
    \label{eq: mapping}
\end{equation}
These equations also apply to the many body case if the system is non-interacting. At $t = \frac{T}{4} = \frac{1}{4}\frac{2\pi}{\omega}$ the rotation is equivalent to a Fourier transform scaled by $m\omega$, which maps $\hat{x}$ to $\hat{p}$, and vice versa. Therefore, two subsequent rotations in traps with frequencies $\omega_1$ and $\omega_2$ lead to a magnification of the matterwave by a factor of $M=\frac{\omega_1}{\omega_2}$. The magnification scheme is illustrated in Fig. 1.  

\begin{figure*}
    \centering
    \includegraphics{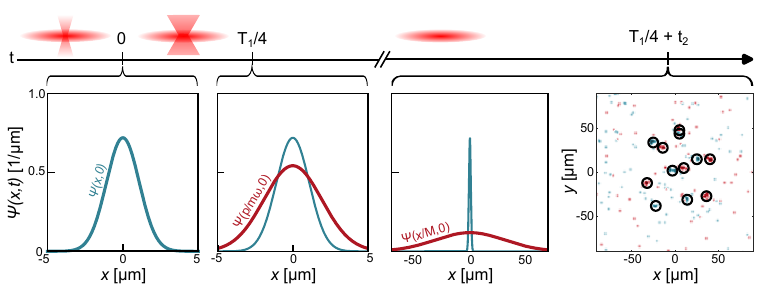}
    \caption{\textbf{Matterwave magnification.} An initial micro meter sized quantum state -- released from an optical tweezer --  evolves non-interacting in the magnifier trap, designed to be approximately harmonic at the length scales of interest. After a quarter trap period ($T_1/4$), the wave function has evolved into its Fourier transform, and we obtain $\Psi(x,T_1/4) = \Psi(p/m\omega_1, 0)$. Subsequently, the wave function is transformed back to real space by a time of flight in the weak radial potential of a light sheet trap (2D-OT),
    resulting in a wave function -- $\Psi(x,T_1/4+t_2) = \Psi(x/M, 0)$ --  magnified by a factor $M$  allowing the spatially resolved imaging of single atoms as shown on the right. The wave function at each time step is shown in red, the initial wave function is shown in blue for reference. The sequence of optical traps is illustrated by the symbols above the timeline.}
    \label{fig: Matterwave magnification}
\end{figure*}

A successful implementation of such a magnification scheme requires that aberrations of the matterwave optics system are well controlled. In practice, this is achieved by making sure that the spatial extent of the wave function $\delta_\text{x}(t)= \sqrt{\braket{\hat{x}(t)^2}}$ never probes the part of the potential where significant deviations from a harmonic shape occur.

A possible implementation is the Gaussian potential created by a laser beam, 
\begin{equation}
U_\text{G}=\tilde{\alpha}I(x)=\tilde{\alpha}\frac{2P}{\pi W^2}\exp{\left(\frac{-2x^2}{W^2}\right)}, 
\end{equation}
where $\tilde{\alpha} = 1/(2\epsilon_0c)\Re(\alpha)$ is proportional to the real part of the polarizability $\alpha$, $P$ is the power of the laser beam, and $W$ is the beam waist in the atom plane. From the series expansion of this potential
\begin{equation}
 U_\text{G}(x) = U_\text{G}(0)\left(1-2\frac{x^2}{W^2}+2\frac{x^4}{W^4}-\frac{4}{3}\frac{x^6}{W^6} + \mathcal{O}\left(x^{8}\right)\right)
\end{equation}
it is evident that
\begin{equation}
   \delta_\text{x}(t) \ll W 
\label{eq: spatial extent}
\end{equation}
is required. At $t=0$ this constraint defines the \emph{field of view} of the matterwave magnifier.
During the rotation in phase space, the initial momentum distribution $\delta_\text{p}(0)$ is converted into a real space distribution $\delta_\text{x}(T_1/4)=\delta_\text{p}(0) / m\omega_1$ at $t=T_1/4$. Through Heisenberg's uncertainty the smallest structure sizes in the initial system $\sigma_\text{x}$ set a lower limit for the momentum distribution, $\delta_\text{p}(0) \geq \hbar/2\sigma_\text{x}$. Therefore, the constraint through~\eqref{eq: spatial extent} at $T_1/4$ defines the \emph{diffraction limit} of the matterwave magnifier. The diffraction limit depends on the waist of the Gaussian potential, and on the available laser power~\cite{supp,Note1}:
\begin{equation}
   \sigma_\text{x} \gg \frac{\sqrt{\pi} \hbar W}{4\sqrt{2|\tilde{\alpha}| P m} }.
   \label{eq: diffraction}
\end{equation}
Note that substantial laser power is required to achieve a diffraction limit much smaller than the field-of-view.


\revision{For the second rotation in phase space, the trap frequency $\omega_2$ is selected such that the magnified smallest structural feature, $M \cdot \sigma_x$, surpasses the optical resolution $\delta_{\mathrm{psf}}$ of the imaging system. Simultaneously, the magnified wave function must remain within the field of view of the optical imaging setup. 
 }


\revision{A mapping of the magnified wave function to the initial wave function is possible for times \( t_2 \) away from \( T_2/4 \), when the contribution of the initial momentum is negligible (i.e. when $\cos(\omega_2 t_2) \delta_p(0)/m \omega_1$ is much smaller than the imaging resolution). In this case, the position operator at time \( T_1/4 + t_2 \) can be approximated as}
\begin{equation}
    \begin{split}
        \hat{x}(T_1/4+t_2) &\approx  \hat{p}(T_1/4) \frac{1}{m \omega_2} \sin(\omega_2 t_2) \\
                          &= -\hat{x}(0) \frac{\omega_1}{\omega_2} \sin(\omega_2 t_2).
    \end{split}
    \label{eq: simplifcation}
\end{equation}
This results in a magnification factor 
\begin{equation}
    M = \frac{\omega_1}{\omega_2} \sin(\omega_2 t_2).
    \label{eq: magnification}
\end{equation}

Our experimental platform is a mesoscopic system of fermionic $^6$Li atoms in two hyperfine states (denoted as spin up and down) in the ground state of a  potential created by two optical traps \cite{supp}. A light sheet (2D-OT) confines the atoms along the vertical (z-) direction - rendering the system quasi-two dimensional (2D).  A second trap confines the atoms in the radial (r-) plane, resulting in a combined potential (OT) with trap frequencies $(\omega_\text{OT,r},\omega_\text{OT,z})/2\pi = \qtylist[list-pair-separator = {, },list-units = bracket]{1420(2);7432(3)}{\hertz}$ (radial and vertical direction, respectively). Due to the small number of atoms, the characteristic size of our system  is set by the harmonic oscillator length,  $l_\text{HO,(r,z)} = \sqrt{\hbar / m \omega_\text{OT,(r,z)} }  \approx \qtylist[list-pair-separator = {, },list-units = bracket]{1.1;0.48} {\micro m}$ in radial and vertical direction, respectively.

To detect the individual atoms, we utilize a free space, single atom and spin resolved fluorescence imaging technique~\cite{bergschneider2018spin}, where each image represents the projection of the wavefunction on N positions, where N is the number of atoms. The random walk of the atoms during fluorescence, caused by the photon recoil~\cite{su2024fast}, sets the rms width of the point spread function of a single atom to  $\delta_\text{psf} = \SI{3.96(5)}{\micro \meter}$ - preventing us from resolving the system without further magnification.

We thus magnify the initial system using a matterwave microscope implemented using two Gaussian potentials -- tailored for accurate imaging down to the smallest structure sizes of interest. As the magnification sequence begins, the wave function is released from the initial radial confinement - the vertical confinement, provided by the 2D-OT stays on during the entire matterwave magnification. In case of interacting systems, we quench off the interactions at the start of the magnification protocol, simultaneously with the release from the radial confinement. The ability to quench interactions at a time scale much faster than any kinematics is vital for the magnification of interacting systems. Here, the quench is realized through a spin flip to a third, essentially non-interacting hyperfine state~\cite{supp,holten2022cooper}. 

The first rotation in phase space is performed by letting the wavefunction expand in a Gaussian potential for a quarter time period $T_1/4$. The potential is created using a focused laser beam  -- referred to as magnifier trap (MT) -- with trap frequency $\omega_1/2\pi = \SI{1130(10)}{\hertz} $, power $P = \SI{4}{W}$ and waist radius $W\approx \SI{35}{\mu m}$ -- chosen to be much larger than the initial extent of the wavefunction. Following \eqref{eq: spatial extent}, the field of view is constrained to be much smaller than $W=\SI{20}{\micro \meter}$ and  condition \eqref{eq: diffraction} sets the smallest resolvable structure sizes to $\sigma_\text{x} \gg \SI{20}{\nano \meter}$. 

\revision{Subsequently, we perform the second rotation in phase space by an expansion in a weak radial potential $U_2$.  This potential is a combination of the radial potential of the light sheet $U_{\text{2D-OT}}$ with $\omega_\text{2D-OT,r}/2\pi = 16.9(1)\,\text{Hz}$, and \revision{a radial waist of $W_\text{2D-OT,r} \approx \SI{600}{\micro \meter}$} and the potential created by the magnetic field curvature $U_\text{B}$, with $ \omega_\text{B,r}/2\pi= \SI{10.9(1)}{\hertz}$. The magnetic potential is either repulsive for the low-field seeking state or attractive for the high-field seeking state, such that the total radial frequency $\omega_2/2\pi = 20.2(3),\,[12.9(3)]\,\text{Hz}$ for \mbox{high-[low-]field seekers}.
This configuration of potentials yields a magnification factor of \mbox{$M = \omega_1/\omega_2 = \num{56(1)}[\num{87(1)}]$}, depending on the specific conditions. For our typical initial system size ($\approx \SI{4}{\micro \meter}$) the magnified wave function remains well within the field of view of the optical imaging setup ($r_\text{FOV} = \SI{350}{\micro \meter}$) and the waist of the second trap $W_\text{2D-OT,r}$.} The smallest structure size of interest in the 2D limit, $l_\text{HO,z} = \SI{480}{\nano \meter} $ is magnified to \mbox{$M\cdot l_\text{HO,z} \approx \SI{26}{\micro \meter}[\SI{42}{\micro \meter}]$}, far surpassing $\delta_\text{psf}$.    

\begin{figure}
    \centering
    \includegraphics{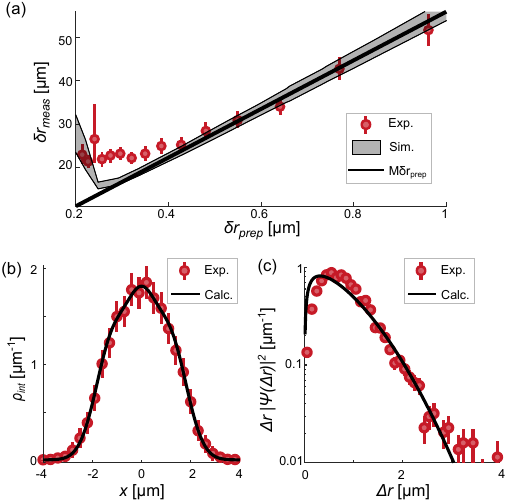}
    \caption{\textbf{Performance of the matterwave magnification} a) 
    We vary the initial extent of the wavefunction $\delta r_\text{prep}$ of a single atom in the ground state of the optical tweezer (OT) and determine the position spread $\delta r_\text{meas}$ from repeated measurements of the atom position after magnification. The black line shows an aberration free magnification with $M = \num{56}$. The experimental results deviate by a factor of $\sqrt{2}$ from this optimum at $\delta r_\text{prep} \approx \SI{300}{\nano \meter}$ -- setting the resolution.  The grey band shows the result of the simulated propagation. b) Measured (red dots) and calculated (black line) real space density integrated along y direction for 6 non-interacting fermions in the ground state of the OT. c) Normalized occurrence of measured (red dots) and calculated (black line)  distances $\Delta r$ in a system of 2 particles with a binding energy $E_\text{B}/h= \SI{1.5}{\kilo \hertz}$ in the ground state of a harmonic trap.  All error bars show the $\SI{95}{\percent}$ confidence interval, determined using a boot-strapping technique.}
    
    \label{fig: performance}
\end{figure}

We characterize the resolution of the matterwave microscope using a single atom in the ground state of the OT as a test target -- shown in Fig 2a. By varying the radial trap frequency of the OT, we vary the \revision{initial system size} $\delta r_\text{prep}$ and determine the root mean square (rms) width after magnification ($\delta r_\text{meas}$) from many experimental implementations of the same quantum state. Our test target -- the harmonic oscillator ground state -- exactly matches the Heisenberg uncertainty limit, and is therefore ideal to probe the diffraction limit described above. In an aberration free system, one obtains $\delta r_\text{meas} = M \cdot \delta r_\text{prep}$ for all $\delta r_\text{meas}$. We observe a deviation from this optimum when the condition in \eqref{eq: diffraction} is not fulfilled and the matterwave probes the anharmonic regions of the MT. We define the resolution as the point where $\delta r_\text{meas} > \sqrt{2} M \delta r_\text{prep}$ - resulting in $\delta r_\text{res} = \SI{300}{\nano \meter}$. Additionally, we perform a numerical simulation of the phase space trajectories in Gaussian traps~\cite{supp}, which matches the experimental measurements, except for very small $\delta r_\text{prep}$.

\begin{figure}
    \centering
    \includegraphics{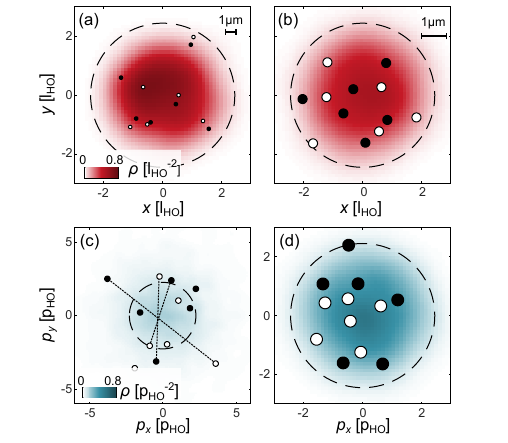}
    \caption{\textbf{Fermionic systems in real and momentum space.} We image a system of 6 spin up and 6 spin down atoms (black/white dots) in real (a,b) and momentum (c,d) space, both in the strongly interacting (a,c) and the non-interacting regime (b,d). The two dimensional histograms show the density distribution, obtained from averaging over many experimental realizations of the same quantum state. The black and white dots show a single, randomly chosen snapshot of the wave-function. The size of the dots represents the resolution. The dashed circles show the Thomas Fermi radius (a,b) and the Fermi momentum (c,d) calculated from the non-interacting system. The lines connecting atoms of opposite spin and momentum in c) serve as a guide to the eye to highlight the opposite momentum correlations.
    }
    \label{fig: realandmomentum}
\end{figure}

We verify the ability of our matterwave microscope to image multiple atoms by comparing the measured and the analytically calculated density of a system of six spin up and six spin down (6+6) non-interacting atoms prepared in the ground state of the OT with radial trap frequency $\omega_\text{OT,r}/2\pi = \SI{1420(2)}{\hertz}$. We magnify the system and obtain the density distribution from averaging over about 1000 experimental realizations of the same quantum state. The measured density integrated over one spatial direction compared to the calculated integrated density is shown in Figure 2b. The measured density is in very good agreement with the calculated density at all positions, demonstrating that this measurement is not limited by the field of view of the matterwave magnifier. The field of view is thus larger than $\SI{4}{\micro \meter}$.

To demonstrate that our microscope can be used to image interacting systems we prepare two weakly-bound atoms in the ground state of the optical tweezer. We calculate the relative wave function of two interacting particles in a harmonic oscillator potential according to \cite{Idziaszek_2006}, again allowing us to compare our measurements to a known target. Using a Feshbach resonance~\cite{Zuern_2013} we set the binding energy~\cite{Idziaszek_2006} to  $E_\text{B}/h = \SI{1.5}{\kilo \hertz}$, with an associated binding length $r_\text{B}= \sqrt{\hbar^2/2mE_\text{B}} \sim \SI{750}{\nano \meter}$.  We obtain a histogram -- shown in Fig 2c -- of relative distances of the two particles from many experimental implementations. The experimental measurements and the theoretical expectation are in good agreement, except at very small distances where we are limited by the resolution of the matterwave magnifier.

The matterwave magnifier can be used to explore a strongly interacting many body system. We prepare 6+6 atoms and set the binding length to $r_\text{B}= \SI{750}{\nano \meter}$, as in the calibration above. To observe molecules we increase the mean inter-particle spacing to $1/\sqrt{n} = \sqrt{4\pi\hbar^2/2mE_\text{F}} \sim \SI{4.5}{\micro \meter}$, by reducing the radial trap frequency to $\omega_\text{OT,r}/2\pi =\SI{180}{\hertz}$. This sets the Fermi energy $E_\text{F}$ -- approximated using the peak density $n$ of the non-interacting system -- to $E_\text{F}/h= \SI{540}{ \hertz}$. 

In Fig. 3a, we show a randomly chosen snapshot of the magnified wavefunction. Here, each spin up atom appears paired up with a spin down atom. The average density - obtained from many experimental implementations - is depicted by the red background. The cloud is smaller than in the non-interacting regime (shown in Figure 3b), where the system size is comparable to the Thomas Fermi radius $r_\text{F} = \sqrt{2E_\text{F}/m\omega_\text{OT,r}^2}$. Note that a different radial trap frequency was used in the non-interacting regime. The position and momenta are thus scaled to the natural units $l_\text{HO,r}$ and $p_\text{HO} = \sqrt{\hbar m\omega_\text{OT,r}}$ of the initial trap. As this alters the harmonic oscillator length $l_\text{HO,r}$, the effective resolution in terms of $l_\text{HO,r}$ is smaller in 3a) than in 3b).


The increase in real space density is accompanied by a decrease in momentum space density as seen in the comparison of Fig 3c) (interacting) and Fig 3d) (non-interacting). The maximum momentum density at a given real space density is governed by Fermi pressure. Beyond that, the momentum density is further decreased by the relative momentum of the pair constituents. A more in-depth analysis of the correlations in momentum space can be found in \cite{holten2022cooper}. 
The measurements in momentum space were obtained by performing a single rotation in phase space in the weak radial potential $\omega_2$~\cite{holten2022cooper}



We have demonstrated matterwave microscopy of a few atom quantum state with single atom and spin resolution. This technique allows for the exploration of correlations of arbitrary order in regimes where the length scales of interest are obfuscated by the resolution of the detection technique. We validate the performance by imaging known systems and show the versatility by imaging 6+6 strongly-interacting atoms.  

Matterwave microscopy allows the in-depth exploration of fermionic pairing with microscopic observables. The technique can also be used to explore more exotic systems such as recently observed~\cite{Chen2024}, weakly bound NaK tetramers. Beyond that, by altering the time in the magnification trap, the matterwave magnification scheme can be extended to measure in arbitrary bases between real and momentum space. This unlocks the possibility of quantum tomography~\cite{Gross2011, Brown2023} and entanglement characterization~\cite{bergschneider2019entanglement, Kitagawa_2011} of many body quantum states.



\providecommand{\noopsort}[1]{}\providecommand{\singleletter}[1]{#1}%
%


\subsection*{Acknowledgments}

This work has been funded by the DFG (German Research Foundation) – Project-ID 273811115 – 
SFB 1225 ISOQUANT, the Germany’s Excellence Strategy EXC2181/1-390900948 (Heidelberg Excellence Cluster STRUCTURES) and the European Union’s Horizon 2020 research and innovation program under grant agreements No.~817482 (PASQuanS) and  No.~725636 (ERC QuStA). This work has been partially financed by the Baden-Württemberg Stiftung.

\paragraph*{Competing Interest}
The authors declare no competing interests.

\paragraph*{Correspondence and requests for materials}
should be addressed to S.B. (brandstetter@physi.uni-heidelberg.de)

\newpage
\section*{Supplemental Material}

\footnotetext[1]{Note that choosing a larger trap frequency $\omega \propto \sqrt{P}w^{-2}$ over a larger waist reduces the laser power required, as long as condition \eqref{eq: spatial extent} is fulfilled.}
\footnotetext[2]{Note that a different radial trap frequency was used in the non-interacting regime. The position and momenta are thus scaled to the natural units $l_\text{HO,r}$ and $p_\text{HO} = \sqrt{\hbar m\omega_\text{OT,r}}$ of the initial trap.}


\setcounter{figure}{0}
\renewcommand{\figurename}{Extended Data Figure}
\renewcommand{\theequation}{S\arabic{equation}}



\paragraph{Preparation and interaction switch off}

We start the experimental sequence by laser cooling $^6$Li atoms using a Zeeman slower and a magneto optical trap. From there the atoms are transferred into a red detuned, crossed beam optical dipole trap, where we perform a sequence of radio frequency pulses to obtain a balanced mixture in hyperfine states $\ket{1}$ and $\ket{3}$ of the $^2$S$_\frac{1}{2}$ Lithium ground state manifold (with states $\ket{1}-\ket{6}$ labeled in ascending order of energies). After a short evaporation, we transfer the atoms into a tightly focused optical tweezer, created using light \revision{with a wavelength of \SI{1064}{\nano \meter}}. Making use of the high densities we perform fast evaporation in this optical tweezer, followed by the spilling technique described in~\cite{Serwane_2011} to arrive at $\approx 30$ atoms. Subsequently we perform a continuous crossover to the quasi-2D regime by ramping \revision{on the power of a vertical optical lattice, created by two beams with a wavelength of $\SI{1064}{\nano \meter}$, interfering under an angle,  creating a light-sheet.} Simultaneously we weaken the radial confinement of the optical tweezer. \revision{The low atom number before this transfer allows us to ensure that atoms are only loaded into single layer of the vertical optical lattice  (2D-OT).} Using the spilling technique introduced in~\cite{bayha2020phase}, we deterministically prepare different atom numbers in the ground state of this 2D optical tweezer. A more detailed account of this preparation sequence can be found in~\cite{bayha2020phase}.

The interactions between hyperfine states $\ket{1}$ and $\ket{3}$ are set via the magnetic offset field using the broad Feshbach resonance of $^6$Li~\cite{Zuern_2013}. To perform matterwave magnification of an initially interacting system, we quench off the interactions by transferring atoms from state $\ket{3}$ to $\ket{4}$ using a two-photon Raman pulse. The $\ket{1}-\ket{4}$ mixture is almost non-interacting with a scattering length set by the singlet scattering length $a_\text{s}/a_\text{0} = \num{47(3)}$~\cite{Abraham_1996}. The pulse duration is $\approx \SI{300}{ns}$, fast enough to preserve the correlations in the system~\cite{holten2022cooper}. \revision{As the residual scattering length $a_\text{s}$ is significantly smaller than $l_\text{HO,z}$, the scattering amplitude $f$ can be approximated by \cite{Dalibard_2011, Parish_2015}
\begin{equation}
    f(k) \approx \sqrt{\frac{i}{k}} \frac{a_\text{s}}{l_\text{HO,z}}\text{,}
\end{equation}
resulting in a scattering cross section 
\begin{equation}
    \sigma(k) \approx \frac{2\pi}{k} \left(\frac{a_\text{s}}{l_\text{HO,z}}\right)^2\text{.}
\end{equation}
Since $a_\text{s}/l_\text{HO,z} \ll 1$ the interactions are negligible. Additionally, as the system is magnified, the scattering cross section increases concurrently with the system size. The system is scale invariant and no resonances occur during the magnification process.}

\begin{figure}
    \centering    \includegraphics{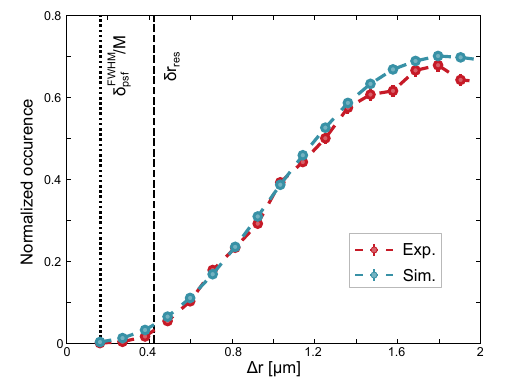}
    \caption{\textbf{Comparison of simulated and measured same spin state distances.} We prepare a non-interacting state of 6+6 atoms and extract a histogram of distances of particles with same spin ($\Delta r$) from many experimental implementations. The results are compared to the results of a Monte-Carlo simulation. The error bars representing the standard error of the mean are smaller than the symbol size.
}
    \label{fig: pauli-hole}
\end{figure}

\paragraph{Diffraction limit}

As discussed above, the diffraction limit is set by the following condition: 
\begin{equation}
   W \gg \delta_x(T/4) = \frac{\delta_p(0)}{m \omega_1} \geq \frac{1}{m \omega_1} \frac{\hbar}{2\sigma_\text{x}}.
\end{equation}
Using the harmonic approximation of the Gaussian potential, the trap frequency $\omega_1$ for atoms of mass $m$ trapped in a Gaussian beam with waist  $W$ and power $P$ is given by 
\begin{equation}
    \omega_1 = \sqrt{\frac{8 P |\tilde{\alpha}|}{m \pi W^4}}.
\end{equation}
This expression allows us to obtain the resolution limit as a function of the Gaussian beam parameters (see Eq. (5))).

\paragraph{Simulation}

The matterwave magnification scheme can be separated into two evolution stages - the T/4 evolution in the magnifier trap (MT) and the time of flight in a potential made from the combination of the radial confinement of the 2D-OT and an additional magnetic trap. We approximate the magnification trap and the 2D-OT with a Gaussian potential and the magnetic trap with a harmonic potential. For the evolution in the magnifier trap this leads to a potential
\begin{equation}
    U_\text{MT} = \frac{m \left(w_\text{MT} \omega_\text{1}\right)^2}{4}\left( 1- \exp \left({\frac{-2x^2}{w_\text{MT}^2}}\right)\right)
\end{equation}
and
\begin{equation}
    \begin{split}
        U_\text{2} &= \frac{m \left(w_\text{r} \omega_\text{2D-OT,r}\right)^2}{4}\left( 1- \text{exp}\left(\frac{-2x^2}{W^2_\text{2D-OT,r}} \right)\right) \\
        &\pm \left( \frac{1}{2} \left(x\omega_\text{B,r}\right)^2m\right)
    \end{split}
\end{equation}
for the combination of the 2D-OT and the magnetic trap. Here $m$ is the mass of the atom, $w_\text{r}$ and $w_\text{MT}$ are the waist of the 2D OT and the magnifier, respectively, $\omega_\text{2D-OT,r}$, $\omega_\text{B,r}$ and $\omega_\text{1}$ are the trap frequencies of the 2D OT, the magnetic trap and the MT.

Using
\begin{equation}
    F = - \frac{\partial U}{\partial x} = m\ddot{x}
\end{equation}
and substituting $\overline{x} = x/w_\text{MT}$ and $\overline{t} = t \cdot 2 \pi/T_1$ where $T_1 = 2\pi/\omega_\text{1}$, gives
\begin{equation}
    \ddot{\overline{x}} = -\overline{x}\left(\overline{t}\right) \exp\left(-2\overline{x}^2\left(\overline{t}\right)\right) 
\end{equation}
for the expansion in magnifier trap. For the second expansion we substitute $\tilde{x} = x/w_\text{r}$ and $\tilde{t} = t \cdot 2 \pi/T_\text{opt}$ where $T_\text{opt} = 2\pi/\omega_\text{2D-OT,r}$ - resulting in 
\begin{equation}
    \ddot{\tilde{x}} = -\tilde{x}\left(\tilde{t}\right) \exp\left(-2\tilde{x}^2\left(\tilde{t}\right)\right)
    + \left(\omega_\text{B,r}/\omega_\text{2D-OT,r}\right)^2
\end{equation}
for the expansion in the combination of the 2D-OT and the magnetic trap. 

We numerically solve these differential equations for different initial conditions, allowing us to compare the final mapping to the ideal case of expansions in two harmonic traps. This was utilized to optimize the parameters of the MT before implementation and to estimate the resolution. For the simulation curve shown in Figure 2, the initial conditions are set by the momentum and real space distribution of a single particle in the ground state of an harmonic trap. For every trap frequency, we sample this initial distribution 10000 times and propagate these initial conditions through the differential equations to obtain $\delta r_\text{sim}$.

\paragraph{Same spin distances}
 
In fermionic systems, Pauli exclusion principle suppresses short distance correlations between atoms of the same spin. The range of these correlations is given by the size of the "Pauli-hole", which scales with $1/E_\text{F}$. 

The ability to resolve two atoms of the same spin state is thus not limited by the diffusion of the atoms during imaging as long as the extent of the magnified Pauli-hole is larger than the full width half maximum of the point spread function of a single atom $\delta_\text{psf}^\text{FWHM} = 2\sqrt{2\ln{2}}\delta_\text{psf} = \SI{9.3(1)}{\micro \meter}$. To verify this, we extract a histogram of same spin state distances $\Delta r$ from experimental measurements of a system of six non-interacting fermions per spin state and compare it to a Monte-Carlo simulation of six non-interacting atoms~\cite{Gajda2016} revealing that the Pauli hole surpasses $\delta_\text{psf}^\text{FWHM}$. Small discrepancies at very short distances stem from the matterwave aberrations.

\paragraph{\revision{Correlation analysis}}

\begin{figure}
    \centering
    \includegraphics{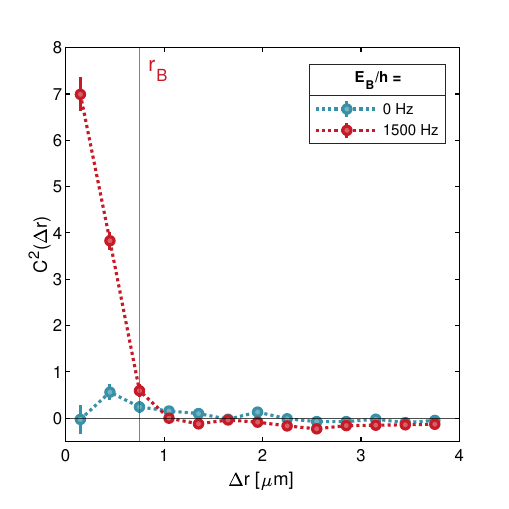}
    \caption{\textbf{Correlation analysis.} The plots show $\mathscr{C}^2(\Delta r)$, the probability of finding two atoms of opposite spin at distance $\Delta r$, normalized by subtracting the contribution of the single particle densities. The blue curve shows the result for the non-interacting system $E_\text{B}/h = \SI{0}{\hertz}$, the red for the strongly interacting system $E_\text{B}/h = \SI{1500}{\hertz}$. The red vertical line marks the binding length $r_\text{B} = \SI{750}{\nano \meter}$ of the strongly interacting system. The error bars represent the standard error of the mean.
}
    \label{fig: C2}
\end{figure}

To verify that the snapshot shown in Figure 3b) is indeed exemplary for the system, we perform an analysis of the correlations in both the interacting and the non-interacting system. The two body correlation function is defined as
\begin{equation}
    \mathscr{C}^{(2)}(\vec{r}_\uparrow,\vec{r}_\downarrow) = \langle n(\vec{r}_\uparrow) n(\vec{r}_\downarrow)\rangle - \langle n(\vec{r}_\uparrow)\rangle\langle n(\vec{r}_\downarrow)\rangle.
\end{equation}
It is convenient to perform the binning in polar coordinates $\vec{r_i} \rightarrow (r_i, \phi_i)$.
As we are only interested in the distance between the particles, we only look at the correlations 

\begin{equation}
    \begin{split}
        &\mathscr{C}^{(2)}(\Delta r) = \int_0^\infty \int_0^\infty \int_0^{2\pi} \int_0^{2\pi} dr_\downarrow dr_\uparrow d\phi_\downarrow d\phi_\uparrow \\ 
        &\mathscr{C}^{(2)} \left(r_\uparrow,\phi_\uparrow,r_\downarrow,\phi_\downarrow\right) 
        \frac{1}{\Delta r}\delta(\Delta r - |(\vec{r}_\uparrow-\vec{r}_\downarrow)|)  r_\downarrow r_\uparrow
    \end{split}
\end{equation}
allowing us to obtain a one-dimensional trap averaged observable, depending only on the opposite spin distance $\Delta r$. The correlations $\mathscr{C}^{(2)}(\Delta r)$ in both the non-interacting  and the strongly interacting regime are shown in Extended Data Figure 2.  We observe the formation of short distance pairs in the strongly interacting system, with a characteristic size set by $r_\text{B}$. As expected, there are no second-order correlations in the non-interacting regime.


\providecommand{\noopsort}[1]{}\providecommand{\singleletter}[1]{#1}%
%


\end{document}